\documentclass[epj]{svjour}
\begin{document}
\title{Relativistic dissipative hydrodynamics
with extended matching conditions
for ultra-relativistic heavy-ion collisions}
\author{Takeshi Osada}

\institute{Department of Physics, Faculty of Liberal Arts and Sciences,
Tokyo City University,\\ Tamazutsumi 1-28-1, Setagaya-ku, Tokyo 158-8557, Japan}
\date{Received: date / Revised version: date}

\abstract{Recently we proposed a novel approach to the formulation
of relativistic dissipative hydrodynamics by extending the
so-called matching conditions in the Eckart frame [Phys. Rev. {\bf
C 85}, (2012) 14906]. We extend this formalism further to the
arbitrary Lorentz frame. We discuss the stability and causality of
solutions of fluid equations which are obtained by applying this
formulation to the Landau frame, which is more relevant to treat
the fluid produced in ultra-relativistic heavy-ion collisions. We
derive equations of motion for a relativistic dissipative fluid
with zero baryon chemical potential and show that linearized
equations obtained from them are stable against small
perturbations. It is found that conditions for a fluid to be
stable against infinitesimal perturbations are equivalent to
imposing restrictions that the sound wave, $c_s$, propagating in
the fluid, must not exceed the speed of light $c$, i.e., $c_s <
c$. This conclusion is equivalent to that obtained in the previous
paper using the Eckart frame [Phys. Rev. {\bf C 85}, (2012)
14906]. }
\titlerunning{Relativistic dissipative hydrodynamics with extended matching conditions}
\maketitle

\section{Introduction}
It is known that one faces problems of instability and violation
of causality in solutions obtained from the relativistic
Naiver-Stokes (NS) equations (observed in the first-order theories
\cite{Eckart1940,Landau1959} which naively extend the
non-relativistic NS equation). Israel and Stewart (IS) provide a
phenomenological framework for a relativistic dissipative fluid
\cite{Israel1976310,Israel1979341} accounting for these problems.
In their model, a possible general form of the non-equilibrium
entropy current is described by the dissipative part of the
energy-momentum tensor and by the particle current up to second
order in the deviation from the equilibrium state. After the
stability and causality of the IS theory have been shown by
Hiscock and Lindblom
\cite{Hiscock1983466,PhysRevD.31.725,PhysRevD.35.3723}, the causal
dissipative hydrodynamical model was adapted to study the dynamics
of hot matter produced in ultra-relativistic heavy-ion collisions
by Muronga \cite{PhysRevLett.88.062302,PhysRevC.69.034903}. In
recent years, the IS model plays an important role in the analysis
of the experimental data obtained by RHIC and LHC (see, for
example, \cite{PhysRevC.84.027902}).

In parallel to the application of IS theory, investigations of the
basis of relativistic dissipative hydrodynamic theory was
continued
\cite{PhysRevC.75.034909,PhysRevD.81.114039,PhysRevLett.105,0954-3899-35-11-115102,0954-3899-36-3-035103,Tsumura2007134,Tsumura2010255,Van:2011yn,springerlink10.1140}.
This is because the IS theory, in its present form, is too general
and complex from the point of view of quantum chromodynamics
(QCD), which is believed to be the fundamental theory for strongly
interacting systems \cite{PhysRevC.75.034909}. Another reason is
that the theory of relativistic dissipative hydrodynamics is not
yet fully understood because, for example, the equation of motion
of the fluid it uses depends on the choice of the Lorentz frame
\cite{Tsumura2007134}  (or on the definition of the hydrodynamical
flow). Since the dissipative part of the energy-momentum tensor
$\delta T^{\mu\nu}$ and the particle current $\delta N^{\mu}$
cannot be determined uniquely by the second law of thermodynamics,
one usually introduces some constraints to fix them. These
constraints are known as matching conditions.

The other reason to introduce these matching (or fitting)
conditions \cite{Israel1979341} is  the necessity of {\it
matching} the energy density and baryonic charge density,
$(\varepsilon, n)$, in a non-equilibrium state to the
corresponding equilibrium densities: ($\varepsilon_{\rm eq}$,
$n_{\rm eq}$), $\varepsilon = \varepsilon_{\rm eq},~n=n_{\rm eq}$,
or, equivalently,
\begin{eqnarray}
  u_{\mu} u_{\nu }\delta T^{\mu\nu}  =0, \quad  u_{\mu} \delta N^{\mu} =0.
\label{eq:standard_matching_condition}
\end{eqnarray}
Matching conditions allow to determine the thermodynamical
pressure, $P_{\rm eq}(\varepsilon_{\rm eq},n_{\rm eq})$, (defined
as work done in isentropic expansion) via the equation of state
for the equilibrium state. Here $P_{\rm eq}$ should be
distinguished from the bulk viscous contribution, $\Pi\equiv
-\frac{1}{3}\Delta_{\mu\nu}\delta T^{\mu\nu}$, present in the
energy-momentum tensor \cite{Israel1979341}. Finally, matching
conditions are also needed because they are necessary for the
thermodynamical stability of the entropy current (see Appendix A
in Ref.\cite{MonnaiPRC80}). However, the matching conditions given
by eq.(\ref{eq:standard_matching_condition}) are not unique. So
far, except of some recent works
\cite{Tsumura2007134,Tsumura2010255}, they were not investigated
in detail.

A state of a relativistic dissipative fluid is described by the
energy-momentum tensor, $T^{\mu\nu}(x)$, and by the baryon number
current, $N^{\mu}$, which obey the conservation laws
\begin{eqnarray}
  && T^{\mu\nu}_{;\mu}=0, \\
  && N^{\mu}_{\mu}=0,
\end{eqnarray}
and the second law of the thermodynamics
\begin{eqnarray}
   S^{\mu}_{;\mu} \ge 0.
\end{eqnarray}

Because of the uncertainty in definition of the flow velocity
$u^{\mu}(x)$ for a non-equilibrium fluid, one needs, unlike in the
case of perfect fluid, to fix the frame for the fluid considered.
Two special frames can be defined: Landau and Eckart.

The Landau frame \cite{Landau1959} is defined by the vanishing of
the energy flow, which consists of heat flow $q^{\mu}$ and net
baryon number flow $V^{\mu}$;
\begin{eqnarray}
 W^{\mu} &\equiv&
  u_{\nu}  T^{\nu\lambda}  \Delta^{\mu}_{\lambda} =0,  \quad
 q^{\mu} =- \frac{\varepsilon_{\rm eq}+P_{\rm eq}}{n_{\rm eq}} ~V^{\mu},
 \label{eq:Landau-frame}
\end{eqnarray}
with $\Delta^{\mu}_{\nu}\equiv g^{\mu}_{\nu} -u^{\mu}u_{\nu}$
being the projection operator orthogonal to the four vector
$u^{\mu}$. In the Eckart frame, the hydrodynamic flow velocity
$u^{\mu}$ (with normalization $u^{\mu}u_{\mu}=1$) is defined by
using baryon charge current $u^{\mu} \equiv
N^{\mu}/\sqrt{N^{\nu}N_{\nu}}$. In this frame one always has
\begin{eqnarray}
 V^{\mu}\equiv \Delta^{\mu}_{\lambda}N^{\lambda}=0, \quad\quad\quad
 q^{\mu}= W^{\mu}. \quad\quad\quad\quad
\end{eqnarray}
Equations of motion for the fluid should not depend on the choice
of Lorentz frame. Since the relativistic dissipative fluid
dynamics with extended matching conditions has been already
formulated in the Eckart frame \cite{OsadaPRC85}, it is
interesting to check it in another Lorentz frame, for example in
the Landau frame. Also, the stability and causality conditions for
a relativistic dissipative fluid should not depend on the Lorentz
frame used. The purpose of this article is therefore to
investigate how the stability and causality conditions should be
imposed on a fluid depending on the Lorentz frame used.

This paper is organized as follows: In Sec.\ref{sec:2}, we
re-formulate relativistic dissipative hydrodynamics with extended
matching condition (as originally introduced in ref.
\cite{OsadaPRC85}) in an arbitrary Lorentz frame with
$V^{\mu}\ne0$ and $W^{\mu}\ne0$ (neither the Eckart frame with
$V^{\mu}=0$ nor the Landau frame with $W^{\mu}=0$). In such a
general frame, the flow velocity field $u^{\mu}(x)$ may be
determined by using coexisting $W^{\mu}$ and $V^{\mu}$. We then
choose the Landau frame and consider a fluid with zero baryon
chemical potential $\mu_b=0$, appearing in the central rapidity
region of the ultra-relativistic heavy-ion collisions. In
Sec.\ref{sec:3}, we check the stability and causality of the fluid
obtained from our model in this frame. We close with
Sec.\ref{sec:4} containing a summary and some further discussion.

\section{Extended matching condition in arbitrary frame}\label{sec:2}
\subsection{General form of off-equilibrium entropy current}

The general off-equilibrium entropy current can be written in the
following simple form using the vector $\phi^{\mu}$ and 2-rank
symmetric tensor $\Phi^{\mu\nu}$,
\begin{eqnarray}
S^{\mu}(x) &\equiv&
 -\alpha \phi^{\mu} +  \beta_{\lambda}\Phi^{\lambda\mu},
\end{eqnarray}
where $\alpha\equiv \mu_b/T$, $\beta^{\mu}=\beta u^{\mu}$ with
$\beta\equiv 1/T$, $T$ and $\mu_b$ are, respectively, the
temperature and baryon chemical potential. In the local
equilibrium case it is given by
\begin{eqnarray}
S^{\mu}_{\rm eq} &\equiv&
 -\alpha\phi^{\mu}_0 +\beta_{\lambda}\Phi^{\lambda\mu}_0,
  \label{eq:equilibrium_entropy_current}\\
  \phi^{\mu}_0 &=&   N^{\mu}_{\rm eq} ,\quad
  \Phi^{\lambda\nu}_0 = T_{\rm eq}^{\lambda\nu}
     - \frac{g^{\lambda\nu}}{3}\Delta_{\alpha\beta}T^{\alpha\beta}_{\rm eq},
     \label{eq:phi0PHI0}
\end{eqnarray}
where $T^{\mu\nu}_{\rm eq}$ and $N^{\mu}_{\rm eq}$ are
equilibrium energy-momentum tensor and baryon charge current:
\begin{eqnarray}
  T^{\mu\nu}_{\rm eq} &=& \varepsilon_{\rm eq} u^{\mu}u^{\nu} -P_{\rm eq}\Delta^{\mu\nu}, \\
  N^{\mu}_{\rm eq} &=& n_{\rm eq} u^{\mu}.
\end{eqnarray}
In this case, the energy-momentum conservation, $T^{\mu\nu}_{{\rm
eq};\mu}=0$, and the baryon number conservation, $N^{\mu}_{{\rm
eq};\mu}=0$, together with thermodynamic relations result in the
locally conserved entropy current:
\begin{eqnarray}
  S^{\mu}_{{\rm eq};\mu} &=& -\alpha_{,\mu}N^{\mu}_{\rm eq}
  +\beta_{\lambda;\mu} T^{\lambda\mu}_{\rm eq} +[\beta^{\mu}P_{\rm eq}]_{;\mu}
  =0. \label{eq:thermodynamical_relation}
\end{eqnarray}
To this current we now introduce dissipative corrections by adding
corresponding dissipative corrections $\delta T^{\mu\nu}$ and
$\delta N^{\mu}$ to the energy-momentum tensor and to the baryon
number current appearing in $\phi_0^{\mu}$ and $\Phi_0^{\mu\nu}$.
In this way one extends the expression of equilibrium entropy
current towards the off-equilibrium entropy current:
\begin{eqnarray}
&& S^{\mu} \equiv -\alpha[\phi^{\mu}_0+\delta \phi^{\mu}]
    +\beta_{\lambda}[\Phi^{\lambda\mu}_0+\delta\Phi^{\lambda\mu}],
     \label{eq:general_entropy1}\\
&& \delta \phi^{\mu} =\delta N^{\mu}, \quad
     \delta \Phi^{\lambda\nu} = \delta  T^{\lambda\nu}
     +\chi ~\frac{g^{\lambda\nu} }{3}\Delta_{\alpha\beta} \delta T^{\alpha\beta}. \quad
     \label{eq:general_entropy2}
\end{eqnarray}
Notice that term proportional to $\Delta_{\alpha\beta} \delta
T^{\alpha\beta}$, which appears due to the natural extension of
eq.(\ref{eq:phi0PHI0}), results in
\begin{eqnarray}
   \lim_{\Pi\to 0} \frac{d}{d\Pi} (u_{\mu}S^{\mu}) ~\ne 0, 
   \label{eq:thermodynamical_instability}
\end{eqnarray}
where $\Pi$ is bulk pressure, cf. Eq. (\ref{eq:delta_Tmunu}) below. 
It means than that entropy density is not maximal in spite of
equilibrium state used, the entropy current
eq.(\ref{eq:general_entropy1}) is thermodynamically unstable
\cite{MonnaiPRC80}. Therefore the term $\Delta_{\alpha\beta}
\delta T^{\alpha\beta}$ is usually dropped (i.e., $\chi$ should be
put equal zero, $\chi\equiv 0$). However, the problem of
thermodynamic instability can be avoided by simultaneously
demanding the natural extension of the form of the entropy current
(eq.(\ref{eq:general_entropy1}) with (\ref{eq:general_entropy2}))
and the following general extension of matching conditions;
\begin{eqnarray}
   \chi\ne 0, \quad
   u_{\mu} u_{\nu} \delta T^{\mu\nu} \ne 0 , \quad
   u_{\mu} \delta N^{\mu} \ne 0.
\end{eqnarray}

\subsection{Extended matching conditions}
To restore thermodynamical stability as discussed above, we
propose to impose the following {\it extended matching conditions}
on the dissipative correction of the energy momentum tensor and
baryon charge current, $\delta T^{\mu\nu}$ and $\delta N^{\mu}$,
respectively:
\begin{eqnarray}
  u_{\mu}\delta T^{\mu\nu}u_{\nu} = \Lambda, \quad
  \delta N^{\mu} u_{\mu} =  \delta n. \label{eq:ext_matching}
\end{eqnarray}
With these conditions, off-equilibrium contributions for the
energy momentum tensor and baryon charge vector in general Lorentz
frame are
\begin{eqnarray}
   \delta T^{\mu\nu} &=& \Lambda u^{\mu}u^{\nu}
   -\Pi \Delta^{\mu\nu} +W^{\mu}u^{\nu} + W^{\nu}u^{\mu} + \pi^{\mu\nu},
   \label{eq:delta_Tmunu}\quad \\
   \delta N^{\mu} &=&\delta n u^{\mu} + V^{\lambda}\Delta_{\lambda}^{\mu},
    \label{eq:delta_Nmu}
\end{eqnarray}
where $\Pi$ is bulk pressure, $\pi^{\mu\nu}$ is shear tensor and
$V^{\mu}$ is net flow of the baryonic charge. In this case, the
off-equilibrium entropy current eq.(\ref{eq:general_entropy1}) is
given by
\begin{eqnarray}
 S^{\mu} = S^{\mu}_{\rm eq}
    -\alpha V^{\lambda}\Delta_{\lambda}^{\mu}
    +\beta  W^{\mu}
    -\beta[ \mu_{\rm b} \delta n  - \Lambda +\chi \Pi] u^{\mu}. \nonumber \\
     \label{eq:entropy_general_expression}
\end{eqnarray}
To ensure thermodynamic stability, one may impose a condition on
the entropy current $S^{\mu}$,
eq.(\ref{eq:entropy_general_expression}), demanding that
\begin{eqnarray}
   \frac{d}{d\Pi} (u_{\mu}S^{\mu})=0.
\end{eqnarray}
This requirement can be satisfied by the following {\it unique}
condition,
\begin{eqnarray}
   \chi \Pi = -\mu_{\rm b} \delta n  + \Lambda.
   \label{eq:thermodynamical_stability}
\end{eqnarray}
Note that, when both $\Lambda$ and $\delta n$ are set equal to
zero, $\chi$ should also be zero, as so far considered widely in
the literature. However, once one assumes that $\Lambda\ne0$
and/or $\delta n \ne 0$, there exists a term proportional to
$\chi$. By the extended thermodynamical stability condition,
eq.(\ref{eq:thermodynamical_stability}), the entropy current in
the arbitrary Lorentz frame is
\begin{eqnarray}
  S^{\mu} = S^{\mu}_{\rm eq}
    -\alpha V^{\lambda}\Delta_{\lambda}^{\mu}+\beta  W^{\mu}.
  \label{eq:exact_1st-order_theory}
\end{eqnarray}
One can obtain the entropy current corresponding to Eckart's or
Landau's formulation in the limit $V^{\mu}\to 0$ and $W^{\mu}\to
0$, respectively. The entropy current
eq.(\ref{eq:exact_1st-order_theory}) can be rewritten using eqs.
(\ref{eq:delta_Tmunu}), (\ref{eq:delta_Nmu}) and thermodynamical
stability condition eq.(\ref{eq:thermodynamical_stability}),
\begin{eqnarray}
  S^{\mu}  &=& S^{\mu}_{\rm eq}
    -\alpha \delta N^{\mu}
    +\beta_{\lambda}  \delta T^{\lambda\mu}
    -\chi \Pi  u^{\mu}. \label{eq:covariant_entropy_current}
\end{eqnarray}
Thus the entropy production in this case is given by
\begin{eqnarray}
 S^{\mu}_{;\mu} = -\alpha_{,\mu} \delta N^{\mu}
                     +\beta_{\lambda;\mu} \delta T^{\lambda\mu}
                     - [\chi \Pi \beta^{\mu} ]_{;\mu} .  \label{eq:PRC85eq19}
\end{eqnarray}
It should be noted here that eq.(\ref{eq:PRC85eq19}) is exactly
the same as eq.(19) in Ref. \cite{OsadaPRC85} obtained in the
Eckart frame when using extended matching condition. This is
because $S^{\mu}_{;\mu}$ is a scaler and so it does not depend on
Lorentz frame. This can be seen in the form of the entropy
current, eq.(\ref{eq:covariant_entropy_current}). The expression
for $S^{\mu}$ has the same form both in Eckart frame ($V^{\mu}=0$
but $\delta N^{\mu}\ne 0$) and in Landau frame ($W^{\mu}=0$ but
$\beta_{\lambda}\delta T^{\lambda\mu}\ne 0$).

The entropy production is explicitly given by
\begin{eqnarray}
  S^{\mu}_{;\mu} &=&
  -(\nabla_{\lambda}\alpha)V^{\lambda}
  +(\nabla_{\lambda}\beta )W^{\lambda}
  +\beta \frac{d u_{\lambda}}{d\tau} W^{\lambda} \nonumber \\
 &-&\beta~ \Pi \theta + \beta ~\nabla_{\langle \mu} u_{\lambda \rangle} \pi^{\lambda\mu}
 \nonumber \\
 &+& [\alpha \frac{d\delta n}{d\tau}- \beta\frac{d\Lambda}{d\tau}]
     +[\alpha\delta n -\beta\Lambda]\theta,
\end{eqnarray}
where $\theta$ is the divergence of the flow velocity field,
$\theta\equiv u^{\mu}_{;\mu}$. Using the definition that
$\alpha=\mu_b \beta$ one can also write the entropy production as
\begin{eqnarray}
  TS^{\mu}_{;\mu} &=&
  [\frac{\nabla_{\mu}\mu_b}{\mu_b}-\frac{\nabla_{\mu}T}{T}]
  [-\mu_b V^{\mu}]
\nonumber \\
&+&
  [~\frac{du_{\mu}}{d\tau}~-\frac{\nabla_{\mu}T}{T}]  ~W^{\mu} \nonumber \\
 &-& \Pi \theta + \nabla_{\langle \mu} u_{\lambda \rangle} \pi^{\lambda\mu}
 \nonumber \\
&+&[\mu_b \frac{d\delta n}{d\tau}-\frac{d\Lambda}{d\tau}]
     +[\mu_b \delta n - \Lambda]\theta.
\end{eqnarray}
Note that the above equation can also be expressed in the
following form:
\begin{eqnarray}
   TS^{\mu}_{;\mu} &=&
 [~\frac{du_{\mu}}{d\tau}~-\frac{\nabla_{\mu}T}{T}] ~\tilde W^{\mu}
\nonumber \\
&+&
  [~\frac{du_{\mu}}{d\tau}~- \frac{\nabla_{\mu}\mu_b}{\mu_b}] ~\tilde V^{\mu}
\nonumber \\
 &-& \Pi \theta + \nabla_{\langle \mu} u_{\lambda \rangle} \pi^{\lambda\mu}
 \nonumber \\
&+&[\mu_b \frac{d\delta n}{d\tau}-\frac{d\Lambda}{d\tau}]
     +[\mu_b \delta n - \Lambda]\theta,
\end{eqnarray}
where $\tilde W^{\mu}\equiv W^{\mu}-\mu_b V^{\mu}$
and $\tilde V^{\mu} \equiv \mu_b V^{\mu}$. \\

\subsection{Constitutive equations for a dissipative fluid in the Landau frame}

We shall now consider the Landau frame ($W^{\mu}\equiv 0$) which
is more relevant in the context of ultra-relativistic heavy-ion
collisions. In particular, we consider a fluid in the central
rapidity region where it is expected that the baryon chemical
potential $\mu_b$ is small. In this paper, we assume that, for
simplicity, $\mu_b=0$ (this implies that $n_{\rm eq}=0$.) In this
case, the entropy production takes the simple form;
\begin{eqnarray}
   TS^{\mu}_{;\mu}
 &=& -\Pi \theta + \nabla_{\langle \mu} u_{\lambda \rangle} \pi^{\lambda\mu}
-[\frac{d\Lambda}{d\tau}+\Lambda\theta]  \nonumber \\
 &=& -\Pi \theta + \nabla_{\langle \mu} u_{\lambda \rangle} \pi^{\lambda\mu}
 -[\frac{d(\chi\Pi)}{d\tau}+(\chi\Pi)\theta] .
\label{eq:entropy_production_with_chi}
\end{eqnarray}
In the second line of eq. (\ref{eq:entropy_production_with_chi})
we have used eq.(\ref{eq:thermodynamical_stability}) with
$\mu_b=0$. For the $\chi$ in
eq.(\ref{eq:entropy_production_with_chi}) we then use
\begin{eqnarray}
   \chi = \kappa + \xi\Pi
   +\xi''\frac{\pi^{\mu\nu}\pi_{\mu\nu}}{\Pi},
   \label{eq:production_central_region}
\end{eqnarray}
i.e., the form of $\chi$ used in eq.(21) in the ref.
\cite{OsadaPRC85} with $W^{\mu}\equiv 0$. The second law of
thermodynamics is guaranteed (with $\zeta$ and $\eta$ being,
respectively, the bulk pressure and shear viscosity, which are all
positive constants) if the entropy production is given by
\begin{eqnarray}
  TS^{\mu}_{;\mu} = \frac{\Pi^2}{\zeta} + \frac{\pi^{\mu\nu}\pi_{\mu\nu}}{2\eta}.
\end{eqnarray}
This requirement determines the following constitutive equation
for,  respectively, bulk and shear pressure:
\begin{eqnarray}
  \frac{\Pi}{\zeta} &=& -(1+\chi) \theta -(\frac{\kappa}{\Pi}+2\xi)\frac{d\Pi}{d\tau},
  \label{eq:constitutive_for_PI} \\
  \frac{\pi_{\mu\nu}}{2\eta} &=& \nabla_{\langle\mu} u_{\nu\rangle}
  -\xi'' \frac{d\pi_{\mu\nu}}{d\tau}.
  \label{eq:constitutive_for_pi}
\end{eqnarray}
Because eq.(\ref{eq:constitutive_for_PI}) includes term
proportional to $1/\Pi~d\Pi/d\tau$, one can introduce into the
bulk pressure $\Pi$ an arbitrary constant $z$ (with dimension
[GeV]$^{4}$) and write
\begin{eqnarray}
  \frac{\Pi}{z} + 2\zeta\xi \frac{d}{d\tau}\frac{\Pi}{z}
 +\frac{\kappa\zeta}{z}\frac{d}{d\tau} \ln \frac{\Pi}{z}
 = -\frac{\zeta}{z} (1+\chi)\theta.
\end{eqnarray}
We shall now consider small perturbations of $\Pi$ fields. The
bulk pressure $\Pi$ can be written as
\begin{eqnarray}
   \Pi = \Pi_0 + \delta \Pi
\end{eqnarray}
with the background reference field $\Pi_0$ and its perturbation
field $\delta \Pi$. One can also regard $\Pi_0$ as the value of
$\Pi$ at initial proper time $\tau_0$, $\Pi_0 = \Pi(\tau_0)$.
Correspondingly, one can write $\theta_0=u^{\mu}_{0;\mu}$,
obtained from the initial flow vector field $u^{\mu}_0$ at
$\tau_0$. In this case, the perturbation field $\delta\Pi$ can be
interpreted as $\delta\Pi = \Pi(\tau)-\Pi(\tau_0)$. In this sense,
$\Pi_0$ is a kind of parameter showing the degree of
non-equilibrium at initial stage. Identifying the arbitrary
constant $z$ with $\Pi_0$ and noticing that $\frac{d}{d\tau}\ln
(1+\frac{\delta\Pi}{\Pi}) \approx \frac{1}{\Pi_0}
\frac{d}{d\tau}\delta \Pi $, one can rewrite the above equation
as:
\begin{eqnarray}
  \Pi_0 &=& -(1+\chi) \zeta \theta_0, \\
 \tau_{\Pi} \frac{d\delta \Pi}{d\tau} +\delta \Pi  &=&  -\zeta(1+\chi)\delta\theta,
\end{eqnarray}
where the relaxation time $\tau_{\Pi}$ is given by
\begin{eqnarray}
 \tau_{\Pi} = \zeta(2\xi + \kappa/\Pi_0 )
\end{eqnarray}
and $\delta\theta=\theta-\theta_0$. It is interesting to note that
$\Pi_0$ contributes to the relaxation time $\tau_{\Pi}$. This
means that relaxation processes may depend on the initial
condition. Note also that if $\kappa=0$ then contribution
$\kappa/\Pi_0$ in the relaxation time would disappear.

A similar approach can also be applied to the $\pi^{\mu\nu}$
field, with perturbation of the shear viscosity around
$\pi^{\mu\nu}= 0$, leading to
\begin{eqnarray}
 \tau_{\pi} \frac{d\delta\pi^{\mu\nu}}{d\tau} + \delta \pi^{\mu\nu} =
 2\eta \nabla^{\langle \mu} u^{\nu\rangle},
\end{eqnarray}
where the relaxation time $\tau_{\pi}$ is given by
\begin{eqnarray}
  \tau_{\pi} = 2\xi'' \eta.
\end{eqnarray}

\section{Stability and causality of the fluid}\label{sec:3}

The stability of a general class of dissipative relativistic fluid
theories was investigated by Hiscock and Lindblom
\cite{Hiscock1983466,PhysRevD.31.725,PhysRevD.35.3723}. Denoting
by $\delta V(x)$ the difference between the actual non-equilibrium
value of a field $V(x)$ and the value  in the background reference
state, $V_0 (x)$, we assume that variations $\delta V$ are small
enough so that their evolution is adequately described by the
linearized equations of motion  describing the background state.
We shall now investigate the stability of the fluid obtained in
our model following a prescription proposed in Ref.
\cite{Hiscock1983466,PhysRevD.31.725,PhysRevD.35.3723}. In what
follows:
\begin{enumerate}
\item

The background reference state
is assumed to be homogeneous in space.
Notice that, unlike in Ref. \cite{Hiscock1983466,PhysRevD.31.725,PhysRevD.35.3723},
in our case it is not an equilibrium state but rather a non-equilibrium one with  $\Pi=\Pi_0$
and with $\pi^{\mu\nu}=0$.
Furthermore, the background space-time is assumed to be flat Minkowski space,
so that all background field variables have vanishing gradients.
\item
We consider following plane wave form of perturbation propagating in $x$ direction
\begin{eqnarray}
 \delta V = \delta V_0 \exp(ikx +\Gamma \tau). \label{eq:plane-wave}
\end{eqnarray}
\end{enumerate}
Linearized equations for dissipative fluid dynamical model are given by
\begin{eqnarray}
\delta [T^{\mu\nu} ]_{;\mu}&=&0,
\end{eqnarray}
with the perturbed energy-momentum tensor:
\begin{eqnarray}
 \delta [T^{\mu\nu}]
 &=&
(\varepsilon_{\rm eq}^* + P_{\rm eq}^*) (\delta u^{\mu}u^{\nu} + u^{\mu}\delta u^{\nu})
+ (\delta \varepsilon_{\rm eq}^* +\kappa \delta \Pi) u^{\mu}u^{\nu}
\nonumber \\
&-&(\delta P_{\rm eq}^* +\delta \Pi) \Delta^{\mu\nu} +\delta \pi^{\mu\nu}. \label{eq:linearized_DeltaTmunu}
\end{eqnarray}
Here $\varepsilon_{\rm eq}^*$ and $P_{\rm eq}^{*}$ are energy density and pressure in the background non-equilibrium state
\begin{eqnarray*}
 \varepsilon_{\rm eq}^* \equiv \varepsilon_{\rm eq} +\kappa\Pi_0 , \quad
  P_{\rm eq}^{*}\equiv P_{\rm eq}+\Pi_0. \quad
\end{eqnarray*}
However,  since $\delta[\Pi_0]=0$
(it has vanishing gradient and is constant in $\tau$),
terms proportional to $\Pi_0$ do not contribute
to the linearized equations eq.(\ref{eq:linearized_DeltaTmunu})
(
we ignore terms like $\Pi_0 \delta u^{\mu}$ in the linearized
equation). Hence, one can replace in the
eq.(\ref{eq:linearized_DeltaTmunu}) $\varepsilon_{\rm eq}^*$ and
$P^*_{\rm eq}$ by the, respectively, $\varepsilon_{\rm eq}$ and
$P_{\rm eq}$. For baryon charge current, since we deal with a
fluid in the region where the baryon chemical potential can be
considered $\mu_b=0$ and the net baryon density $n_{\rm eq}=0$,
one has $\delta [N^{\mu} ] \equiv 0$.

The perturbed fluid dynamical fields must satisfy constraints
\begin{eqnarray*}
u_{\mu}\delta u^{\mu}=0, \quad \delta \pi^{\mu\nu} u_{\mu}=0.
\end{eqnarray*}
Hence, in the rest frame of fluid, $u^{\mu}=(1,0,0,0)$, the proper
time $\tau$ component of the flow velocity field vanishes, $\delta
u^{\tau}\equiv 0$. We therefore obtain the following linearized
equations for the energy-momentum tensor and baryon number
current:
\begin{eqnarray}
  \delta [T^{\mu\nu}]_{;\mu}
  &=& (\varepsilon_{\rm eq} +  P_{\rm eq})((ik\delta u^{x}) u^{\nu} +\Gamma \delta u^{\nu}) \nonumber \\
  &+& (\Gamma \delta \varepsilon_{\rm eq}+\kappa\Gamma\delta \Pi ) u^{\nu}
    - (\nabla^{\nu} \delta P_{\rm eq}+\nabla^{\nu}\delta \Pi) \nonumber \\
  &+& (ik)\delta\pi^{x\nu} =0.
\end{eqnarray}
The linearized constitutive equations for $\Pi$ and $\pi^{\mu\nu}$
have the form (with  $\tilde\kappa\equiv 1+\kappa$ ) :
\begin{eqnarray}
\frac{(1+\tau_{\Pi}\Gamma )}{\zeta} \delta \Pi
&=& -\tilde\kappa (ik)\delta u^{x}, \label{eq:Linearized_Pi} \\
\frac{(1+\tau_{\pi}\Gamma )}{2\eta} \delta \pi^{\mu\nu}
 &=&-\frac{1}{2}(ik)[ \delta^{\mu}_{x}\delta u^{\nu}
 +\delta^{\nu}_x\delta u^{\mu} -\frac{2}{3} \delta^{\mu\nu}\delta u^x]. \nonumber \\
\end{eqnarray}
The parameters $\xi$ and $\xi''$ introduced above have been
absorbed in the expressions for relaxation time, $\tau_{\Pi}$ and
$\tau_{\pi}$, respectively. On the other hand, the parameter
$\kappa$ in the expression of the entropy production is kept and
not absorbed in $\tau_{\Pi}$. Its role will be discussed later.

All perturbation equations can be expressed in concise matrix form:
\begin{eqnarray}
  M^A_B \delta Y^B=0, \label{eq:Linearized Matrix}
\end{eqnarray}
where $\delta Y^B$ represents the list of fields.
The system matrix $M^A_B$ can be expressed in a block-diagonal form when one
chooses the following set of perturbation variables \cite{PhysRevD.31.725}
\begin{eqnarray}
  \delta Y^B &=&\{~
  \delta \varepsilon_{\rm eq}, \delta u^x, \delta \Pi, \delta \pi^{xx}, \nonumber \\
&& ~\delta u^y,\delta \pi^{xy}, ~\delta u^z, \delta \pi^{xz}, 
\delta \pi^{yz}, \delta \pi^{yy}-\delta \pi^{zz} \}.\quad
\end{eqnarray}
In this case,
\begin{eqnarray}
  {\bf M}= \left( \begin{array}{cccc}
  {\bf Q} & & \\
 & {\bf R} & & \\
 & & {\bf R} &  \\
 & & & {\bf I} \\
\end{array}   \right) ,
\end{eqnarray}
where the matrices ${\bf Q}$ and ${\bf R}$ are given by
\begin{eqnarray}
{\bf Q} &=&
\left(
\begin{array}{cccc}
    \Gamma & ikh_{\rm eq} & \kappa\Gamma & 0 \\
    ik\frac{\partial P_{\rm eq}}{\partial \varepsilon_{\rm eq}}&
    \Gamma h_{\rm eq}
    & ik & ik \\
    0 & ik \tilde\kappa & \frac{1+\tau_{\Pi}\Gamma}{\zeta} & 0 \\
    0 & ik& 0 & \frac{1+\tau_{\pi}\Gamma}{4\eta/3} \\
\end{array}
\right), \nonumber \\
   {\bf R}&=& \left(
   \begin{array}{cc}
    h_{\rm eq} \Gamma  & ik \\
    ik & \frac{1+\tau_{\pi}\Gamma}{\eta} \\
  \end{array} \right),
\end{eqnarray}
respectively, and {\bf I} is the $2\times2$ unit matrix. The $h_{\rm eq}$ denotes the enthalpy density
which is defined by $h_{\rm eq} \equiv \varepsilon_{\rm eq}+P_{\rm eq}$.
For $\Gamma$ and $k$ satisfying dispersion relation
\begin{eqnarray}
[{\rm det}{\bf M}]=[{\rm det}~{\bf R}]^2[{\rm det}~{\bf Q}]=0,
\end{eqnarray}
one has plane-wave solution such as eq.(\ref{eq:plane-wave}) for
the linearized equations of the system eq.(\ref{eq:Linearized
Matrix}).

In what follows we shall discuss in detail the stability of
transverse and longitudinal modes separately.

\subsection{Propagation of the transverse mode}
The dispersion relation obtained by setting
\begin{eqnarray}
 \eta~ {\rm det} ({\bf R}) =
 (\tau_{\pi} h_{\rm eq}) \Gamma^2 + h_{\rm eq} \Gamma + \eta k^2 =0
\label{eq:transverse_mode}
\end{eqnarray}
corresponds to the solution of the perturbation
equation which is referred to as the so-called transverse mode.
The solution of the above equation (\ref{eq:transverse_mode}) is given by
\begin{eqnarray}
  \Gamma =\frac{-h_{\rm eq}\pm \sqrt{
  h_{\rm eq}^2 -4\eta (\tau_{\pi}h_{\rm eq}) k^2}}{2(\tau_{\pi} h_{\rm eq})}.
\end{eqnarray}
Note that we have always ${\rm Re} [\Gamma] <0$ independent of the
value of $k$, which means that any small perturbation propagating
in the transverse direction (perpendicular to the $x$ axis,
direction which the perturbation wave propagates) will be damped
with time $\tau$. Since the general solution is given by a linear
combination of those solutions of the transverse mode,  one can
say that the plane-wave solution of the mode is stable against
small perturbation.

Note also that when wave number $k \ge k_c$, where
\begin{eqnarray}
   k_c = \sqrt{\frac{h_{\rm eq}}{4\eta\tau_{\pi}}},
\end{eqnarray}
the linear perturbation wave propagates towards the transverse
direction, but waves with wave number $k<k_c$ are damped.

\subsection{Propagation of the longitudinal mode}
Frequencies of the so-called longitudinal
mode (propagating parallel to the $x$ direction)
are given by the roots of the following dispersion relation:
\begin{eqnarray}
 \left[\frac{4\eta}{3\tau_{\pi}}\frac{\zeta}{\tau_{\Pi}} \right] {\rm det}({\bf Q})
 \equiv \sum_{n=0}^{n=4} q_n \Gamma^n =0,
 \label{eq:longitudinal_mode}
\end{eqnarray}
where the coefficients $q_n$ are given by
\begin{eqnarray}
   q_4 &=& h_{\rm eq}, \\
   q_3 &=& h_{\rm eq} \left( \frac{1}{\tau_{\Pi}}+\frac{1}{\tau_{\pi}} \right), \\
   q_2 &=& h_{\rm eq} \left( \frac{1}{\tau_{\Pi}\tau_{\pi}}
   +\frac{\partial P_{\rm eq}}{\partial \varepsilon_{\rm eq}} k^2 \right) \nonumber \\
   &&\quad + \left( \frac{4\eta}{3\tau_{\pi}} +\tilde\kappa (1-\kappa
  \frac{\partial P_{\rm eq}}{\partial \varepsilon_{\rm eq}})
  \frac{\zeta}{\tau_{\Pi}} \right) k^2, \\
  q_1 &=& h_{\rm eq} \left( \frac{1}{\tau_{\Pi}}+\frac{1}{\tau_{\pi}} \right)
  \frac{\partial P_{\rm eq}}{\partial \varepsilon_{\rm eq}}
  k^2 \nonumber \\
&&\quad + \left( \frac{4\eta}{3\tau_{\pi}}\frac{1}{\tau_{\Pi}}
  +\tilde\kappa (1-\kappa
  \frac{\partial P_{\rm eq}}{\partial \varepsilon_{\rm eq}})
  \frac{\zeta}{\tau_{\Pi}}\frac{1}{\tau_{\pi}} \right) k^2,  \\
  q_0 &=& h_{\rm eq} \frac{\partial P_{\rm eq}}{\partial \varepsilon_{\rm eq}}
  \frac{1}{\tau_{\Pi}}\frac{1}{\tau_{\pi}} k^2.
\end{eqnarray}
When all coefficients $q_n$ of the fourth-order equation
(\ref{eq:longitudinal_mode}) have the same sign, the four (complex
or real) solutions of the real part is definitely negative. In
this case, the general solution which is a linear combination of
those four solutions, is stable. Since $q_4$, $q_3$, and $q_0$
are positive defined, then the stability condition sought after is
that $q_2$ and $q_1$ must be simultaneously positive. The
condition is
\begin{eqnarray}
    1-\kappa
    \frac{\partial P_{\rm eq}}{\partial \varepsilon_{\rm eq}}
    \bigg|_{n_{\rm eq}=0}
    >0. \label{eq:stability_1}
\end{eqnarray}
Using thermodyanmical relation $\frac{\partial P_{\rm
eq}}{\partial \varepsilon_{\rm eq}}\Big|_{n_{\rm eq}}= c_s^2
+\frac{1}{T}\frac{\partial P_{\rm eq}}{\partial s_{\rm eq}}
\Big|_{\varepsilon_{\rm eq}}$, where $c_s^2\equiv \frac{\partial
P_{\rm eq}}{\partial \varepsilon_{\rm eq}}\Big|_{s_{\rm eq}} $ is
the adiabatic velocity of sound, one can rewrite the stability
condition eq.(\ref{eq:stability_1}) with the following
\begin{eqnarray}
 c_s^2 + \frac{1}{T}\frac{\partial P_{\rm eq}}{\partial s_{\rm eq}}
\Big|_{\varepsilon_{\rm eq}} < \frac{1}{\kappa},
 \label{eq:stability_2}
\end{eqnarray}
which is exactly the same condition found in previous work
\cite{OsadaPRC85} in the Eckart frame. Note that, when $\kappa \to
0$, one finds that the speed of sound can exceed unity violating
causality. On the other hand, when $\kappa$ is a finite number
restricted by
\begin{eqnarray}
 \frac{1}{T}\frac{\partial P_{\rm eq}}{\partial s_{\rm eq}}\Big|_{\varepsilon_{\rm eq}}
  \le \frac{1}{\kappa} \le 1+
 \frac{1}{T}\frac{\partial P_{\rm eq}}{\partial s_{\rm eq}}\Big|_{\varepsilon_{\rm eq}},
 \label{eq:causality}
\end{eqnarray}
then the velocity of sound satisfies $0\le c_s \le 1$. Thus, when
the condition for $\kappa$, eq.(\ref{eq:causality}) is satisfied,
the fluid is stable against small perturbations and evolves
without violating causality.

\section{Summary and concluding remarks}\label{sec:4}

We have proposed a novel formulation of the relativistic
dissipative hydrodynamical model in an arbitrary frame by using
extended matching conditions
\begin{eqnarray*}
 u_{\mu} u_{\nu} \delta T^{\mu\nu} =\Lambda, \quad u_{\mu} \delta N^{\mu} =\delta n.
\end{eqnarray*}
To apply the above extended matching conditions, we have also
generalized the form of the entropy current for non-equilibrium
state [cf. eq.(\ref{eq:entropy_general_expression})]~:
\begin{eqnarray*}
   S^{\mu} = S^{\mu}_{\rm eq}
    -\alpha V^{\lambda}\Delta_{\lambda}^{\mu}
    +\beta  W^{\mu}
    -\beta[ \mu_{\rm b} \delta n  + \Lambda -\chi \Pi] u^{\mu}.
\end{eqnarray*}
The phenomenological parameter $\chi$ introduced in the
generalization of the entropy current can be fixed by the extended
thermodynamic stability condition,
\begin{eqnarray*}
  \mu_{\rm b} \delta n  + \Lambda -\chi \Pi =0.
\end{eqnarray*}
(Note that in the usual formulation $\chi\equiv 0$, because of
$\Lambda=0$ and $\delta n=0$). Taking the thermodynamical
stability condition into account, the entropy current is given by
\begin{eqnarray*}
  S^{\mu} &=& S^{\mu}_{\rm eq}
    -\alpha V^{\lambda}\Delta_{\lambda}^{\mu}
    +\beta  W^{\mu} \nonumber \\
  &=&   S^{\mu}_{\rm eq}
    -\alpha \delta N^{\mu}
    +\beta_{\lambda}  \delta T^{\lambda\mu}
    -\chi \Pi  u^{\mu}.
\end{eqnarray*}
As seen in the above equation, the last term $\chi\Pi u^{\mu}$ is
the new correction term. The corresponding entropy production
evidently does not depend on the Lorentz frame considered. It is
given by
\begin{eqnarray*}
  S^{\mu}_{;\mu} = -\alpha_{,\mu} \delta N^{\mu}
  + \beta_{\lambda;\mu} \delta T^{\lambda\mu}
  - [\chi\Pi\beta]_{;\mu} .
\end{eqnarray*}
In this paper, we chose the Landau frame and considered a
dissipative fluid with zero chemical potential $\mu_b\equiv 0$,
for simplicity. For this case, the $\chi$ was assumed as
\begin{eqnarray*}
  \chi = \kappa + \xi \Pi + \xi'' \frac{\pi^{\mu\nu}\pi_{\mu\nu}}{\Pi}.
\end{eqnarray*}
The physical meaning of $\kappa$, $\xi$ and $\xi''$ are revealed
in the discussion on the stability and causality of the fluid in
Sec.\ref{sec:4}. The $\kappa$ is related to the bound for the
speed of sound wave of the fluid and $\xi$ and $\xi''$ are related
to the relaxation of the off-equilibrium system. In the linearized
field equations, the speed of sound $c_s$ is actually restricted
so that $0 \le c^2_s \le 1$ when $\kappa$ is chosen in the
following range:
\begin{eqnarray*}
  \frac{1}{T}\frac{\partial P_{\rm eq}}{\partial s_{\rm eq}}\Big|_{\varepsilon_{\rm eq}}
  \le \frac{1}{\kappa} \le 1+
 \frac{1}{T}\frac{\partial P_{\rm eq}}{\partial s_{\rm eq}}\Big|_{\varepsilon_{\rm eq}}.
\end{eqnarray*}
On the other hand, the relaxation time of the perturbed $\Pi$ and
$\pi^{\mu\nu}$ fields are respectively given by
\begin{eqnarray*}
   \tau_{\Pi} = (2\xi +\kappa/\Pi_0)\zeta, \quad \tau_{\pi} = 2\xi''\eta.
\end{eqnarray*}
The parameter $\kappa$ also contributes to the relaxation time and
may bring in a contribution to the initial condition characterized
by $\Pi_0$.

We therefore conclude that, when the matching conditions
\begin{eqnarray*}
 \Lambda = \chi \Pi =
 \kappa \Pi + \xi \Pi^2  + \xi'' \pi^{\mu\nu}\pi_{\mu\nu}
\end{eqnarray*}
and $\delta n=0$ (also $\mu_b=0$) are imposed, the relativistic
dissipative fluid can be applied to an analysis of the phenomena
observed in the ultra-relativistic heavy-ion collisions not only
in the Eckart frame, as discussed in \cite{OsadaPRC85}, but also
in the Landau frame. The conditions that should be imposed seem to
be independent of the Lorentz frame used, i.e.,
\begin{eqnarray}
 \frac{\lambda T}{\tau_w} \le h_{\rm eq} \quad\mbox{and}\quad
  c_s^2 \le \frac{1}{\kappa} -
  \frac{1}{T}\frac{\partial P_{\rm eq}}{\partial s_{\rm eq}}\Big|_{\varepsilon_{\rm eq}},
\end{eqnarray}
where $\lambda$ and $\tau_{w}$ are thermal energy
conductivity and relaxation time of the thermal energy
conduction, respectively \cite{OsadaPRC85}. In the Landau frame,
$\frac{\lambda T}{\tau_w}$ should be regarded as 0 because of
the definition of the frame, $W^{\mu}=0.$

\begin{acknowledgement}
We gratefully acknowledge discussions with T. Koide. The author
would like to warmly thank dr E. Infeld for reading this
manuscript.

\end{acknowledgement}

\end{document}